\documentclass[aps,twocolumn,prl,unsortedaddress]{revtex4}%
\usepackage{amsmath}
\usepackage{graphicx}%
\usepackage{amsfonts}%
\usepackage{amssymb}
\begin{document}

\title{New Model and Numerical Test of a High Temperature Pairing Mechanism in Stripes}
\author{Oron Zachar}
\affiliation{UCLA Dept. of Physics, Los Angeles, CA 90095, USA.}

\begin{abstract}
We introduce a new model and mechanism of high temperature pairing in stripes.
We propose a way to unambiguously test it by numerical simulations. For
example, the implementation of our mechanism in a 6-leg t-J ladder model has
the effect of making the spin gap of the doped 6-leg ladder be about 4 times
bigger than the spin gap of the same ladder at half filling.

\end{abstract}
\maketitle

The cuperate superconductors manifest a remarkable variety of unusual and
unexplained phenomena. Let us focus on superconductivity. As highlighted in
their name, the high temperature superconductors are distinguished by a high
pairing energy scale $\Delta_{s}\approx35meV$. The most significant missing
element in our understanding of superconductivity in the cuperates is the
microscopic pairing mechanism. Known microscopic characteristics, such as the
strength of spin exchange interactions, $J\approx1500K$, restrict candidate mechanisms.

Therefore, from a theoretical point of view we ask what models can, in
principle, produce such high pairing energies over the range of doping as
observed in the cuperates? Our aim in this paper is to provide such a model
based on stripes. We also propose a straight foreword numerical test of our
proposed mechanism. The model, and the main effect which is in operation, is
distinct from previous proposals.

The paper is organized as follows: First, we present our model and propose a
numerical test to verify or negate the operation of the core effect at work
(we call it an ''environment spin decoupling effect''). Then, we use a
two-chain model to illustrate the environment spin decoupling effect using
tractable analytical methods. We conclude with a discussion of the
implications of our model within the general phenomenology of stripes and
superconductivity in the cuperates.

\section{\textsl{Stripe model and numerical test}}

Experimentally, the spin gap in cuperate superconductors is nearly independent
of doping $\delta$ in the range $0.06<\delta<0.15$ (with a magnitude
$\Delta_{s}\equiv2\Delta\left(  0\right)  \approx800K\sim J/2$
\cite{PseudoGap-review-Timusk99}). In the same doping range, the distance
between stripes changes by a factor two. Therefore, any mechanism of pairing
via stripes must be independent of the distance between stripes. The
difficulty of achieving such a mechanism can be appreciated by thinking of
stripes as a periodic arrangement of ladders:

The mechanism of pairing in stripes is supposed to operate on each stripe
independently. Therefore, the same mechanism should manifest itself in a
single ladder model where the number of legs of the ladder is determined by
the stripe period. Since the stripe period change from 4 to 8 lattice spacings
between doping $\delta=0.12$ and $\delta=0.06$, it implies that a mechanism of
pairing in stripes should be such that the pairing of mobile fermions in the
corresponding 4-leg and 8-leg ladder models be the same!! Indeed, the model
and mechanism which we propose is aiming to achieve exactly that.%

\begin{figure}
[ptb]
\begin{center}
\includegraphics[
height=1.5835in,
width=1.6016in
]%
{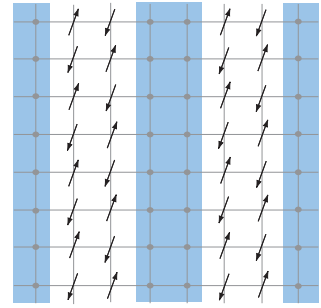}%
\caption{Schematic picture of a stripe state. Darker shaded regions represent
higher hole consentration. Arrows represent magnitude of spin moments on Cu
sites. Periodic charge stripes (hole rich) are separated by nearly undoped
regions. }%
\label{Fig-StripePicture}%
\end{center}
\end{figure}

In order to have a sharp goal in mind, we pose the following theoretical
question: Is it possible to have a spin gap in an even-leg doped ladder which
is several times bigger than the spin gap of the same undoped ladder?
Numerical investigation \cite{Poilblanc02-4leg-Gap} seem to indicate that the
answer is negative on uniform ladders. In this paper, we investigate
non-uniform ladder models.

One idea in this direction is to make the bond interactions inhomogeneous.
Arrigoni \& Kivelson \cite{ArigoniKivelson-4leg-Inhomogeneous} considered a
4-leg Hubbard ladder model where the bare hopping interaction was dimerized
transverse to the ladder (while maintaining the same uniform hopping
interactions along all the legs). They showed that indeed the maximum spin gap
of the doped ladder may increase substantially.

We introduce an inhomogeniety not in the interactions but instead in the
on-site chemical potential. e.g., for t-J type ladders The model is
characterized by three parameters $\left\{  t,J,\mu\right\}  $. In Fig.2 we
draw a 3-leg and 6-leg ladder versions of our models. To begin with, we shall
discuss the ladder model from a purely technical perspective and defer the
discussion of its association with the stripe state to the concluding section
of this paper. The simplest anisotropic ladder model consists of a chemical
potential shift $\mu$ in two neighboring legs of an N-leg ladder.
Consequently, holes are preferentially concentrated in these two legs. Hence,
upon doping by holes away from half filling, the chemical potential difference
leads to an inhomogeneous distribution of the holes. We argue that a
significant renormalization of interactions leads to an effective phase
separation of the system into two incoherent subsystems at intermediate
temperatures $T_{2}^{\ast}<T<T_{1}^{\ast}$ - One is a doped 2-leg t-J ladder,
and the second is an undoped $\left(  N-2\right)  $ leg ladder. This
decoupling effect is the main effect which leads to a high energy pairing
scale. It is a direct consequence of the inhomogeniety of the hole
distribution in the ladder due to the confining potential whose strength is
characterized by $\mu$. At lower temperatures $T<T_{2}^{\ast}$ coherence is
established by the proximity effect. As we discuss in section-IV, this lower
temperature coherence is related to the superconducting $T_{c}$.%

\begin{figure}
[ptb]
\begin{center}
\includegraphics[
height=3.371in,
width=3.1496in
]%
{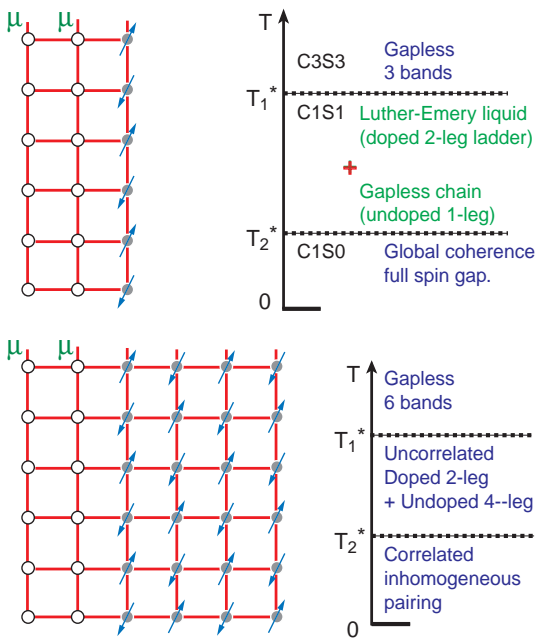}%
\caption{The empty circles denote some incommensurate electron filling
fraction. The other legs are half-filled or nearly half-filled and manifest
significant local antiferromagnetic correlations. In this figure we draw the
examples of anisotropic 3-leg and 6-leg ladders, and the associated phase
diagram at moderate doping. }%
\label{Fig-LadderModels}%
\end{center}
\end{figure}

First we focus on zero temperature properties which can be checked by
numerical simulations and serve as a qualitative and quantitative test of our
claimed principle mechanism. Consider the example of a 6-leg ladder (see
Fig.2). Upon doping, the system separates into a doped 2-leg ladder and an
undoped 4-leg ladder. i.e., the minimal spin gap in the system is increased to
the order of a 4-leg ladder spin gap $\Delta_{s}\approx J/6$ (i.e., a five
fold increase compared with the undoped 6-leg ladder gap $\Delta_{s}\approx
J/20$!). This can be check by numerical simulation methods such as DMRG and
CORE \cite{Poilblanc02-4leg-Gap}. An additional conjectured property of our
model is the following: The pairing energy of the holes is that of a two leg
ladder, and largely independent of the spin correlations in the remaining
undoped legs of the ladder. This should also be amenable to numerical experiments.

\section{\textsl{1D model with competing intra- band and inter- band
interactions}}

The above proposed numerical experiments can quantitatively test our proposed
model, yet, in the following section we would like to develop the reader's
analytical intuition about the competition and renormalization of interactions
in the model. Instead of directly thinking of the dynamic decoupling of a
ladder system into two incoherent sub-systems, let us think of the reverse
route: Consider two 1D systems - one is a doped 2-leg Hubbard ladder and the
second is an undoped or lightly doped N-leg Hubbard ladder. Turning on hopping
interaction $t_{\perp}$ between the two ladders, there is a competition
between the intra-ladder interaction which maintains the spin gap (and
pairing) on the 2-leg ladder and inter-ladder interactions which form
coherence across the whole coupled system. Similar issues of
confinement/deconfinment have risen in the context of coupling
\emph{equivalent ladder} systems where the \emph{tuning parameter is
inter-ladder interactions} (e.g., $t_{\perp}\neq t$ or $J_{\perp}\neq J$). In
contrast, our model starts with \emph{uniform} \emph{bare interactions
}($t_{\perp}=t$, $J_{\perp}=J$)\emph{ }while inhomogeniety of doping induced
by\emph{ tuning an inhomogeneous chemical potential}. We argue that similar
confinement/deconfinment competition of intra- and inter- ladder interactions
is at play. We would like to demonstrates this competition of interactions and
the core effect where a spin gap and pairing is established on a sub-part of
the system and it's spin degrees of freedom decouple from the rest of the
system which may remain gapless. Such an effect is unique to models of coupled
\emph{inequivalent} 1D systems. Below we examine the simplest model of such type.

Consider a model of two \emph{inequivalent} 1D chains (or two inequivalent 1D
bands): Band-A has a Fermi wave number $k_{A}$ and attractive interactions
$-U_{A}$. Band-B has a different Fermi wave number $k_{B}$ and repulsive
interactions $+U_{B}$. The filling fractions are such that $k_{A}<k_{B}%
\leq\frac{\pi}{2}$. Hence, in the absence of inter-band interactions, the
groundstate of band-A is a Luther-Emery (LE) liquid characterized by a spin
gap (due to the attractive $-U_{A}$ interactions) while band-B has gapless
spin excitations. If band-B is half filled, $k_{B}=\frac{\pi}{2}$, then it is
equivalent to a Heisenberg spin chain, or otherwise if $k_{B}<\frac{\pi}{2}$
then band-B is a Luttinger liquid. Qualitatively, band-A and band-B correspond
to the doped 2-legs and the remaining $N-2$ legs of the inhomogeneous ladder
model.( The fact that in a 2-leg ladder the spin gap is due to repulsive
interactions while in this two-chain model the spin gap in band-A is due to
attractive interactions is not of conceptual significance).

It is straight forward to determine that the only marginally relevant
\emph{inter-band} interactions (with respect to the non-interacting fixed
point) are inter-band spin exchange $M\vec{s}_{A}\left(  x\right)  \cdot
\vec{s}_{B}\left(  x\right)  $ and Josephson coupling (i.e., inter-band pair
hopping) $J\left[  P_{A}^{\dagger}\left(  x\right)  P_{B}\left(  x\right)
+h.c.\right]  $ (where the j-band pairing operator $P_{j}(x)=\psi_{j,\uparrow
}^{\dagger}\psi_{j,\downarrow}^{\dagger}$). Note, in the case that band-B is
half filled then the inter-band Josephson coupling is perturbatively
irrelevant. We shall discuss both $k_{B}<\frac{\pi}{2}$ and $k_{B}=\frac{\pi
}{2}$ cases.

The Hamiltonian density which describe our one dimensional (1D) two-band model
is
\begin{align}
H\left(  x\right)   &  =H_{0}^{B}\left(  x\right)  +U_{B}\rho^{B}\left(
x\right)  \rho^{B}\left(  x\right) \\
&  +H_{0}^{A}\left(  x\right)  -U_{A}\rho^{A}\left(  x\right)  \rho^{A}\left(
x\right) \nonumber\\
&  +M\vec{s}_{A}\left(  x\right)  \cdot\vec{s}_{B}\left(  x\right)  +J\left[
P_{A}^{\dagger}\left(  x\right)  P_{B}\left(  x\right)  +h.c.\right] \nonumber
\end{align}
$\rho_{\uparrow}^{A}\left(  x\right)  =\psi_{A,\uparrow}^{\dagger}\left(
x\right)  \psi_{A,\uparrow}^{{}}\left(  x\right)  $, $\vec{s}_{A}\left(
x\right)  =\psi_{A\alpha}^{\dagger}\left(  x\right)  \frac{\mathbf{\sigma
}_{\alpha\beta}}{2}\psi_{A,\beta}^{{}}\left(  x\right)  $. Additional
marginally irrelevant interactions (e.g., inter-band charge density
interactions) which do not interest us here have been dropped.

The three interaction $\left\{  -U_{A},M,J\right\}  $ are all marginally
relevant in the initial stages of renormalization. Yet, the first crucial
observation that we make is that the intra-band attractive interaction
$-U_{A}$ and the inter-band exchange interaction $M$ are competing. i.e.,
$-U_{A}$ and $M$ control distinct fixed points for which in the final stages
of renormalization either $M\rightarrow0$ or $U_{A}\rightarrow0$ respectively.
The fixed points and renormalization group flows are drawn in Fig-4. These
results are derived using the bosonization method as summarized below.%

\begin{figure}
[ptb]
\begin{center}
\includegraphics[
height=2.0358in,
width=2.2554in
]%
{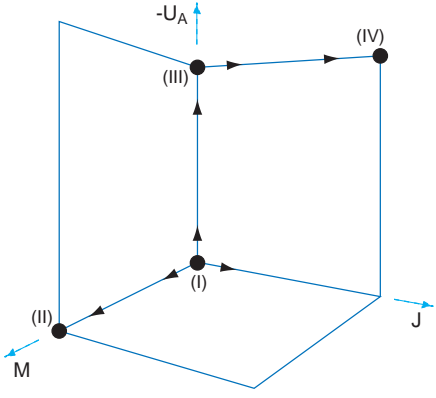}%
\caption{Fixed points and renormalization group flows.}%
\label{Fig-FixedPoints}%
\end{center}
\end{figure}

The fixed points of our model cam be inferred by observation from the
bosonized representation of the model \cite{1D-ref}; Using the bosonized form
of chiral (left and right moving) fermion fields $\psi_{L/R,\sigma}^{\dagger
}\left(  x\right)  =F_{\sigma}\sqrt{\frac{\mu}{2\pi}}:e^{+i\sqrt{\pi}\left[
\theta_{\sigma}\left(  x\right)  \pm\phi_{\sigma}\left(  x\right)  \right]
}:$, where $\theta_{\sigma}(x)=\int_{-\infty}^{x}dx^{\prime}\Pi_{\sigma
}(x^{\prime})$, and $\left[  \Pi_{\sigma}(x^{\prime}),\phi_{\sigma}(x)\right]
=-i{\delta}(x^{\prime}-x)$, $\sigma=\uparrow,\downarrow$. The anticommuting
Klein factors, $\{F_{\sigma},F_{\sigma^{\prime}}\}=\delta_{\sigma
,\sigma^{\prime}}$, are needed for the proper anticommutation of fermions with
different spin. (Normal ordering eliminates the short-distance regularization
factor $1/\sqrt{a}$ that is common in the literature). In the following, all
operators are implicitly assumed to be normal ordered. Also, we absorb the
infra-red cut-off $\mu$ factors into the interaction coefficients). As
commonly done, we re-express the operators in terms of bosonic spin fields
$\phi_{s}(x)=\frac{1}{\sqrt{2}}[\phi_{\uparrow}-\phi_{\downarrow}]$, and
charge fields $\phi_{c}(x)=\frac{1}{\sqrt{2}}[\phi_{\uparrow}+\phi
_{\downarrow}]$, and correspondingly defined momenta $\Pi_{s}$ and $\Pi_{c}$.
e.g., pair creation $P^{\dagger}\sim\frac{1}{\pi}e^{+i\sqrt{2\pi}\theta_{c}%
}\cos\left(  \sqrt{2\pi}\phi_{s}\right)  $ and spin flip operator $s^{+}%
\sim\frac{1}{\pi}e^{+i\sqrt{2\pi}\theta_{s}}\cos\left(  \sqrt{2\pi}\phi
_{s}\right)  $. The resulting form of the model Hamiltonian density is%
\begin{align}
H\left(  x\right)   &  =H_{0}^{A}\left(  x\right)  -\frac{U_{A}}{\pi^{2}}%
\cos\left(  \sqrt{8\pi}\phi_{s}^{A}\right)  \\
&  +H_{0}^{B}\left(  x\right)  +\frac{U_{B}}{\pi^{2}}\cos\left(  \sqrt{8\pi
}\phi_{s}^{B}\right)  \nonumber\\
&  +\frac{M}{\pi^{2}}e^{+i\sqrt{2\pi}\left[  \theta_{s}^{A}-\theta_{s}%
^{B}\right]  }\cos\left(  \sqrt{2\pi}\phi_{s}^{A}\right)  \cos\left(
\sqrt{2\pi}\phi_{s}^{B}\right)  \nonumber\\
&  +\frac{J}{\pi^{2}}e^{+i\sqrt{2\pi}\left[  \theta_{c}^{A}-\theta_{c}%
^{B}\right]  }\cos\left(  \sqrt{2\pi}\phi_{s}^{A}\right)  \cos\left(
\sqrt{2\pi}\phi_{s}^{B}\right)  \nonumber\\
&  +h.c.+\left\{  \text{marginally irrelevant interactions}\right\}  \nonumber
\end{align}
where
\begin{align}
H_{0}^{j}\left(  x\right)   &  =\frac{1}{2}\left[  K_{c}^{j}\left(  \Pi
_{c}^{j}\right)  ^{2}+\frac{1}{K_{c}^{j}}\left(  \partial_{x}\phi_{c}%
^{j}\right)  ^{2}\right]  \\
&  +\frac{1}{2}\left[  K_{s}^{j}\left(  \Pi_{s}^{j}\right)  ^{2}%
+\frac{1}{K_{s}^{j}}\left(  \partial_{x}\phi_{s}^{j}\right)  ^{2}\right]
.\nonumber
\end{align}
The difference between attractive and repulsive intra-band interaction is
expressed by the Luttinger parameters $K_{c}^{A}>1$ and $K_{c}^{B}>1$ for the
charge sector. It is useful to define $\theta_{c}^{\pm}=\frac{1}{\sqrt{2}%
}\left(  \theta_{c}^{B}\pm\theta_{c}^{A}\right)  $ and $\theta_{s}^{\pm
}=\frac{1}{\sqrt{2}}\left(  \theta_{s}^{B}\pm\theta_{s}^{A}\right)  $. For the
purpose of delineating the potential spin gap fixed points, one needs to keep
track only of interaction terms which may develop a non-zero expectation
value. Since by general theorems it always holds that $\left\langle
\cos\left(  \sqrt{4\pi}\theta_{s}^{-}\right)  \cos\left(  \sqrt{4\pi}\phi
_{s}^{-}\right)  \right\rangle =0$, then we need to keep only the $\cos\left(
\sqrt{4\pi}\theta_{s}^{-}\right)  \cos\left(  \sqrt{4\pi}\phi_{s}^{+}\right)
$ part of the exchange interaction. Also, since $K_{c}^{B}>1$ then the $U_{B}$
interaction is irrelevant and can also be dropped. Hence we arrive at
\begin{align}
H &  =H_{0}^{B}+H_{0}^{A}\left(  x\right)  -\frac{U_{A}}{\pi^{2}}\cos\left(
\sqrt{4\pi}\left(  \phi_{s}^{+}-\phi_{s}^{-}\right)  \right)  \\
&  +\frac{2M}{\pi^{2}}\cos\left(  \sqrt{4\pi}\theta_{s}^{-}\right)
\cos\left(  \sqrt{4\pi}\phi_{s}^{+}\right)  \nonumber\\
&  +\frac{2J}{\pi^{2}}\cos\left(  \sqrt{4\pi}\theta_{c}^{-}\right)  \left[
\cos\left(  \sqrt{4\pi}\phi_{s}^{+}\right)  +\cos\left(  \sqrt{4\pi}\phi
_{s}^{-}\right)  \right]  \nonumber\\
&  +\left\{  \text{irrelevant and marginally irrelevant interaction}\right\}
\nonumber
\end{align}
The competition between $M$ and $U_{A}$ is inferred from the fact that at a
fixed point where the inter-band interaction $M$ is relevant then
$\left\langle \cos\left(  \sqrt{4\pi}\theta_{s}^{-}\right)  \right\rangle
\neq0$ and $\left\langle \cos\left(  \sqrt{4\pi}\phi_{s}^{-}\right)
\right\rangle =0$ and hence the intra-band $U_{A}$ interaction is irrelevant.
The resulting phase diagram depicted in Fig.3 shows the fixed points and
renormalization group (RG) flows. In an expanded following publication we
shall elaborate on the complete RG equations and estimates of the spin gap
magnitude renormalization.

As discussed by \cite{1D-Classify}, a fixed point is specified by the gapless
order parameters. The enumeration of gapless modes is straight forward by
observation. Only the $\phi_{c}^{+}$ charge mode remains gapless, as expected
from an over all incommensurately filled system. To distinguish between fixed
points with the same number of gapless modes, it suffice here to focus on the
pairing order parameters%
\begin{align*}
\Delta_{e}  &  =\frac{1}{2}\left[  \left(  R_{B\uparrow}L_{B\downarrow
}+L_{B\uparrow}R_{B\downarrow}\right)  +\left(  R_{A\uparrow}L_{A\downarrow
}+L_{A\uparrow}R_{A\downarrow}\right)  \right] \\
&  \sim e^{-i\sqrt{\pi}\theta_{c}^{+}}\left[  e^{-i\sqrt{\pi}\theta_{c}^{-}%
}\cos\left(  \phi_{s}^{+}\right)  \cos\left(  \phi_{s}^{-}\right)  \right] \\
\Delta_{o}  &  =\frac{1}{2}\left[  \left(  R_{B\uparrow}L_{B\downarrow
}+L_{B\uparrow}R_{B\downarrow}\right)  -\left(  R_{A\uparrow}L_{A\downarrow
}+L_{A\uparrow}R_{A\downarrow}\right)  \right] \\
&  \sim e^{-i\sqrt{\pi}\theta_{c}^{+}}\left[  e^{-i\sqrt{\pi}\theta_{c}^{-}%
}\sin\left(  \sqrt{\pi}\phi_{s}^{+}\right)  \sin\left(  \sqrt{\pi}\phi_{s}%
^{-}\right)  \right] \\
\Delta_{te}  &  =\frac{1}{2}\left[  \left(  R_{B\uparrow}L_{A\downarrow
}+L_{B\uparrow}R_{A\downarrow}\right)  +\left(  R_{A\uparrow}L_{B\downarrow
}+L_{A\uparrow}R_{B\downarrow}\right)  \right] \\
&  \sim e^{-i\sqrt{\pi}\theta_{c}^{+}}\left[  e^{+i\sqrt{\pi}\phi_{c-}}%
\sin\left(  \sqrt{\pi}\phi_{s}^{+}\right)  \sin\left(  \sqrt{\pi}\theta
_{s}^{-}\right)  \right] \\
\Delta_{to}  &  =\frac{1}{2}\left[  \left(  R_{B\uparrow}L_{A\downarrow
}+L_{B\uparrow}R_{A\downarrow}\right)  -\left(  R_{A\uparrow}L_{B\downarrow
}+L_{A\uparrow}R_{B\downarrow}\right)  \right] \\
&  \sim e^{-i\sqrt{\pi}\theta_{c}^{+}}\left[  e^{+i\sqrt{\pi}\phi_{c-}}%
\cos\left(  \sqrt{\pi}\phi_{s}^{+}\right)  \cos\left(  \sqrt{\pi}\theta
_{s}^{-}\right)  \right]
\end{align*}
The subscripts $e/o$ indicate even/odd parity under exchange of band indices.

Table-1 delineate the gapless modes at each fixed point.%

\begin{tabular}
[c]{|c||c|c|c|c|}\hline
$k_{B}<\frac{\pi}{2}$ & FP-I & FP-II & FP-III & FP-IV\\\hline\hline
Dominant interaction &  & $M$ & $-U_{A}$ & $-U_{A},J$\\\hline\hline
Gapless modes & $C2S2$ & $C1S0$ & $C2S1$ & $C1S0$\\\hline
$\Delta_{e}$ & $\surd$ & 0 & $\surd$ & $\surd$\\\hline
$\Delta_{o}$ & $\surd$ & 0 & $\surd$ & $\surd$\\\hline
$\Delta_{te}$ & $\surd$ & $\surd$ & 0 & 0\\\hline
$\Delta_{to}$ & $\surd$ & $\surd$ & 0 & 0\\\hline
\end{tabular}

If band-B is half filled, $k_{B}=\frac{\pi}{2}$, then it is equivalent to a
Heisenberg spin chain \cite{ZacharTsvelik-KondoHeisenberg} and the fixed
points are modified since band-B does not support gapless charge modes and the
Josephson coupling $J$ is perturbatively irrelevant. Hence, modes $\Delta
_{te}$ and $\Delta_{to}$ are eliminated. For an elaborate discussion of FP-II
in this case see \cite{ZacharTsvelik-KondoHeisenberg}.

\section{\textsl{Discussion}}

From a general theoretical perspective, our modified ladder model has the
following special properties: (a) The spin gap of the doped ladder model is
higher than of the half filled ladder. (b) The predominant flow towards fixed
point III (of increased pairing interactions in part of the ladder system
while diminishing spin exchange coupling with the rest of the ladder system)
leads at intermediate temperatures to the establishment of a spin gap on a
sub-part of the ladder (the two hole rich legs), where the temperature scale
and the magnitude of the spin gap are characterized by those of a doped 2-leg
ladder, and are independent of the number of legs of the full ladder. (c)
Consequently, at intermediate temperatures the ladder separates into two
subsystems of which the spin dynamics is respectively decoupled. (d) At low
temperatures, global coherence is established by the proximity effect
(Josephson coupling) as expressed by the flow towards fixed point IV in the
phase diagram.

We here argue that the model captures the core physics of the stripe structure
in doped antiferromagnets. While the mechanism for the creation of stripes is
not generally agreed upon, the end result is manifestly a state of local
charge inhomogeneity of elongated domains of alternating hole rich and hole
poor strips. i.e., alternating strips with different effective local chemical
potential. Therefore, a repeated arrangement of our inhomogeneous ladder model
is a model representation of the stripe structure (where the width of the
ladder corresponds to the charge period of the stripe). Such a coupled ladder
model has the following properties: At intermediate temperatures a spin gap is
established independently on each hole rich stripe, while the hole poor strips
may remain gapless. Residual interaction may lead to coupling between the hole
poor regions across the spin gapped hole rich stripe, but at intermediate
temperatures such coupling may just as well be too weak and leave the hole
poor regions decoupled and thus manifest only local AFM correlations. At low
temperature, global coherence is established by coherent pair tunneling,
leading to a superconducting state.

These characteristics of our model realize the goals of the general framework
first presented in \cite{SpinGapProximity}. Yet, in terms of the pairing
mechanism itself, the core effect presented here is orthogonal to that in
\cite{SpinGapProximity}. The pairing mechanism proposed in
\cite{SpinGapProximity} had the pairing arising out of the \emph{coherent
interaction} between hole rich and hole poor regions. In contrast, the core
mechanism in our new model is a \emph{decoupling} of the dynamics of the hole
rich and hole poor regions. The proximity effect introduced by
\cite{SpinGapProximity} comes into operation in our model only at low
temperatures to establish global coherence (generating the flow towards fixed
point IV). Elsewhere we shall expand on the significant advantages (compared
with previous stripe based proposals) of the model and mechanism introduced
here to address phenomenological properties of pairing and superconductivity
in the cuperates.

\begin{acknowledgments}
We thank Steve Kivelson and Jiang-ping Hu for valuable discussions, and the
support of NSF grant \#4-444025-KI-21420-2.
\end{acknowledgments}

\end{document}